\documentclass[sigconf]{acmart}
\AtBeginDocument{%
  \providecommand\BibTeX{{%
    \normalfont B\kern-0.5em{\scshape i\kern-0.25em b}\kern-0.8em\TeX}}}

\usepackage{multirow}
\usepackage{subcaption}

\begin{document}

\title{Step Out of Your Comfort Zone: More Inclusive Content Recommendation for Networked Systems}


\author{Jiaxin Wu}
\authornote{Both authors contributed equally to this research.}
\email{jiaxinw3@illinois.edu}
\affiliation{%
 \institution{University of Illinois at Urbana-Champaign}
 \city{Champaign}
 \state{Illinois}
 \country{USA}}

\author{Supawit Chockchowwat}
\authornotemark[1]
\email{supawit2@illinois.edu}
\affiliation{%
 \institution{University of Illinois at Urbana-Champaign}
 \city{Champaign}
 \state{Illinois}
 \country{USA}}

\renewcommand{\shortauthors}{Wu and Chockchowwat}

\begin{abstract}
Networked systems are widely applicable in real-world scenarios such as social networks, infrastructure networks, and biological networks. Among those applications, we are interested in social networks due to their complexity and popularity. One crucial task on the social network is to recommend new content based on special characteristics of the graph structure. In this project, we aim to enhance the recommender systems by preventing the recommendations from leaning towards contents from closed communities. To counteract the bias, we will consider information dissemination across network as a metric to assess the recommendation for contents e.g. new connections and news feed. We use academic collaboration network and user-item interaction datasets from Yelp to simulate an environment for connection recommendations and to validate the proposed algorithm.
\end{abstract}

\maketitle

\section{Introduction}
Graphs have been utilized in a wide variety of practical applications, including but not limited to social networks, biology networks, and interconnected large-scale systems. In the applications of social networks, graphs are composed of individuals and their relationships with others, represented as nodes and edges. Similarly, warehouses and routes of a supply chain network can be modeled as a directed graph with nodes and edges. For those graphical systems, especially for social networks, one important application is how to accurately recommend contents or connections for existing users. Profound research has been conducted to establish reliable and efficient recommender systems. For example, \citet{Konstas2009} introduce a collaborative filtering recommendation system. In their work, they utilize network information including personal preferences and underlying users' communities to enhance the performance of the recommender system. \citet{Yang2012} demonstrate that the accuracy of the recommendation system can be increased by learning the category-specific social trust circles from the network data. Moreover, \citet{Fan2019} apply the graphical neural network to predict user-item rating. The model separately learns the user and item latent variables from the two user-user and user-item graphs, and train the predictor on these latent variables. 

All the aforementioned works try to improve the recommender systems' performance by applying additional information from the graphs, such as communities of users, trust circles as well as users' interactions. Nonetheless, such additional information tends to be community-specific; as a consequence, the users would likely connect within their community or heavily receive homogeneous recommendations. The cycle of a user interacting based on recommendations and recommendations constructed from the user's interactions leads to an undesirably disconnected graph. The overall utility of such recommendation is questionable: out-of-community recommendations can be more preferable for users. Furthermore, the dissemination of information should not be limited by specific communities identified. For example, on Youtube, a user may prefer to have coverings for the same subject from different perspectives; or on TikTok, the user may want to discover new contents different from his/her "learned" interests. As a result, in this project, we want to avoid the over-constrained recommender systems by considering information dissemination crossing the entire network via user-item interaction graph.

Information dissemination has been used as a measure in the literature to evaluate the structure of the graph. For instance, \citet{Tong2012Gelling} have demonstrated an algorithm to increase the information dissemination by adding edges to the graph. However, it assumes that all edges have the same addition cost, which is false in the recommendation problem; in particular, target users have a preference for each connection. Generalizing the idea for the social network settings, out-of-community connections should be recommended based on information dissemination metrics together with the likelihood of users' interactions.



Thus, we adopt DiffNet \citet{wu2019neural} to formulate an appropriate optimization model to recommend potential user-item connections based on social network or user-user interactions (Section \ref{sec:diffnet}). We also study information dissemination in existing graphs and propose an integration of dissemination into aforementioned optimization problem to train DiffNet in order to optimize recommendation accuracy and dissemination capability simultaneously (Section \ref{sec:dissemination}). We then empirically study the effect of dissemination factor and draw DiffNet's Pareto front on accuracy-dissemination trade-off (\ref{sec:experiments}). In general, we are able to formulate a recommender algorithm to smoothly incorporate information dissemination as an additional criterion.

\section{Network Diffusion for Recommendation} \label{sec:diffnet}
As for traditional recommendation tasks, many studies have been conducted to find sophisticated user and item embeddings in latent space and thus to directly predict users' preferences based on the embedded representations\citet{Rendle2012,Guo2017,He2017}. However, the performance of the embedding based recommendation technique is usually deteriorated because of the sparsity of the user-item interactions. This drawback motivates the research about how to incorporate external information other than user/item embeddings to enhance the recommendation system. For instance, with the abundant information from the social networks, interactions between the users can be utilized to learn the user's preference towards a specific item even tough the user-item interaction of such a pairwise relation is missing. The intuition is that people who are in close contact may share similar interests and such social network information can be utilized to tackle the data sparsity issue. Thus, how to leverage the underlying information in the social networks becomes a promising direction in the recent literature \citet{Guo2015, Jiang2014, tang2013social}. Among all those social network aided recommendation frameworks, in this study, we first adopt the influence diffusion neural network (DiffNet) proposed by \citet{wu2019neural} to analyze the applicability of using social network information in the recommender system.

The essential component of the DiffNet is to capture the user's interest and to model the propagation process of the user's influence to his/her neighbors k-hops away. Such an iterative interest/influence diffusion process can change the latent user embeddings at each propagation step. And it's beneficial to comprehend the precise user/item embeddings during the diffusion process to improve the recommendation performance. In the following sections, we first briefly summarize the key components of the DiffNet, and demonstrate it's original performance on the real world dataset.

\subsection{DiffNet Architecture}
\begin{figure*}
    \centering
    \includegraphics[width = 0.9\linewidth]{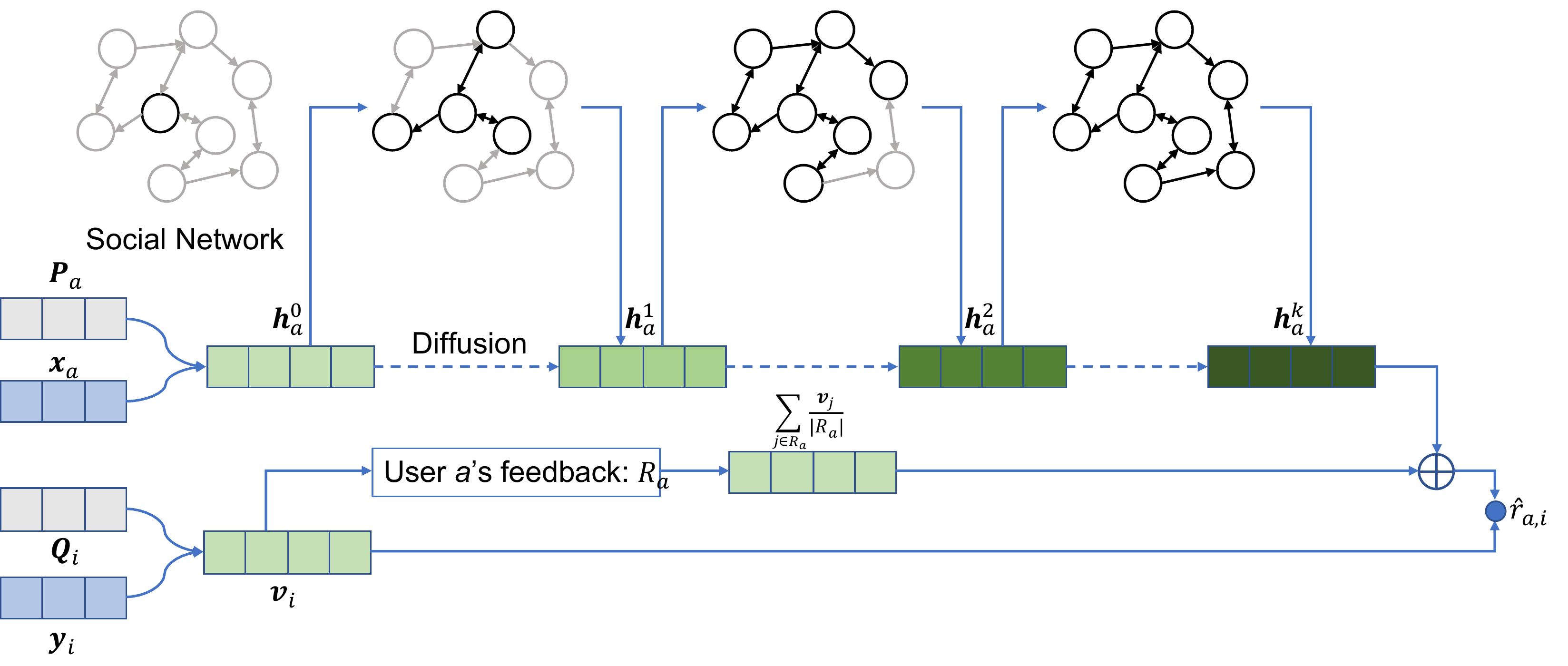}
    \caption{The flowchart of the DiffNet model: recommendations are predicted based on the diffusion results from the auxiliary social network and the user/item embeddings.}
    \label{fig:diffnet_flowchart}
\end{figure*}

There are four major components of the DiffNet: embedding layer, fusion layer, influence diffusion layer, and prediction layer. And the overall flowchart of the DiffNet is depicted in \autoref{fig:diffnet_flowchart}.

\textbf{Embedding layer}: Similar to the collaborative filtering, free embeddings of the user and item can be obtained from e.g. matrix factorization and neural network based filtering. Let $\mathbf{P}\in \mathbb{R}^{M\times D}$ and $\mathbf{Q}\in \mathbb{R}^{N\times D}$ represent the embeddings of the users and items, where $M/N$ is the number of users/items and $D$ is the dimension of the latent space. These two free embeddings are treated as the initial inputs to the DiffNet.

\textbf{Fusion layer}:
Besides the embeddings of the users/items, other useful features can also be used as the input. For instance, by using the Word2vec model, features can be constructed from the embedding of each word in the reviews given to businesses on Yelp, the comments left on posts on Twitter, or the descriptions of images on Instagram. Once the corpus of a given dataset has been analyzed, the feature vector of each user/item can be constructed by averaging all the learned word vectors for the user/item. For the user $a$, let $\mathbf{x}_a$ denote the feature vector. Then the free embedding of the user $\mathbf{P}_a$ can be fused with the $\mathbf{x}_a$ by employing a simple neural network with one fully connected layer:
\begin{equation}
    \mathbf{h}_a^0=g(\mathbf{W}^0\times (\mathbf{x}_a, \mathbf{P}_a)),
\end{equation}
where $\mathbf{W}$ is the weight; $g$ is a nonlinear function e.g. a fully connected layer; $\mathbf{x}_a$ and $\mathbf{P}_a$ have been concatenated first. Similarly, the fusion layer for the item $i$ can be expressed as:
\begin{equation}
    \mathbf{v}_i=g(\mathbf{W}\times (\mathbf{y}_i, \mathbf{Q}_i)).
\end{equation}

\textbf{Influence diffusion layer}: After we obtain the initial fused user embedding $\mathbf{h}_a^0$, then we need to diffuse such a user's influence information in the social network. And the influence diffusion layer models the dynamic diffusion process for the user's latent preference. This process can be represented as a $k$-layer structure: each layer takes the user embedding $\mathbf{h}_a^{k-1}$ from the previous layer and outputs the updated embedding $\mathbf{h}_a^{k}$ as the propagation goes on. And the propagation forwards by one hop at each layer. Specifically, the updated user embedding consists of two parts: the previous embedding and the influence diffusion from the trusted users at current layer. Notice that, in a connected social network, a user $a$ trusts every users if $k$ is large enough. Let $S_a$ denote the set of trusted users of user $a$, then the influence diffusion from trusted users for user $a$ is
\begin{equation}
    \mathbf{h}_{S_a}^{k+1}=Pool(\mathbf{h}_{b}^{k}|b\in S_a),
\end{equation}
where the $pool$ is a pooling operation, e.g. taking the maximum of or averaging the input. Then the overall updated user embedding can be formulated as a nonlinear mapping:
\begin{equation}
    \mathbf{h}_{a}^{k+1}=g(\mathbf{W}^k \times (\mathbf{h}_{S_a}^{k+1}, \mathbf{h}_{a}^{k})),
\end{equation}
where again $g$ is a nonlinear function and $\mathbf{W}^k$ is trainable weights.

\textbf{Prediction layer}: With both the user embedding $\mathbf{h}_{a}^{k}$ after $K$-hops diffusion and the fused item vector $\mathbf{v}_i$ on hand, the final user representation can be obtained from
\begin{equation}
    \mathbf{u}_a=\mathbf{h}_{a}^{K}+\sum_{i\in R_a}\frac{\mathbf{v}_i}{|R_a|},
\end{equation}
where $R_a$ is the set of items that user $a$ has shown interest. This equation states that the latent representation of each user has two parts: the user embedding from the social network diffusion process, and the average item embedding from his/her known preferences in the dataset. This user embedding is more comprehensive by considering both the information of the social network and the user's historical preferences. And in order to make predictions for potential user-item interactions, we can adopt the traditional approach of taking the inner product of the item embedding and the final user representation:
\begin{equation}
    \hat{r}_{ai}=\mathbf{v}_{i}^T\mathbf{u}_a.
\end{equation}
This $\hat{r}_{ai}$ can be interpreted as the predicted probability of having a connection between the user $a$ and the item $i$. And in this way, based on the recommendation prediction results, a new graph $G:=(V:=\{a,i\},E:=a\times i)$ can be established.

\subsection{Numerical Study}
To evaluate the effectiveness of the DiffNet for recommending new contents, we utilize the well-known Yelp dataset \citet{yelp_dataset}. This data set consists of two parts: a social network with 17237 users, 129455 edges and a matrix recording the rating from those users to 37378 businesses. The user/item feature matrix $\mathbf{X}$/$\mathbf{Y}$ is first derived from the Word2vec model by analyzing the reviews. The dimension of the feature is set to be 150 for both users and items. To evaluate the recommendation performance, we utilize two metrics: hit ratio (HR) and normalized discounted cumulative gain (NDCG). Given a top-N ranking list based on the recommendation result $\hat{r}_{ai}$, the HR measures the proportion of the number of recommendations that the user truly likes. And NDCG not only considers the occurrence of an appropriate recommendation but also takes the rank of each recommendation into consideration: the higher the rank of a correct recommendation, the larger the score.

Notice that there are two important parameters that can significantly affect the performance of the DiffNet: the dimension $D$ of the free embedding $\mathbf{P}$/$\mathbf{Q}$ and the top-N values $N$ in the evaluation metrics. Thus in \autoref{tab:yelp_num_result} we show the numerical results of the HR and NDCG for different settings of $D$ and $N$. We include the recommendation performance from the traditional collaborative filtering based technique i.e. SVD++ \citet{guo2015trustsvd} as a comparison. Moreover, comparing to other graph diffusion based approach, e.g. graphical convolutional network (GCN) based PinSage \citet{ying2018graph}, the DiffNet provides better performance in terms of the HR and NDCG: the PinSage achieves a HR score around 0.30 for $D=32$ and $N=10$, while that for NDCG is 0.18, that are both worse than the results of DiffNet.
\begin{table*}[!h]
    \centering
    \begin{tabular}{cccccc}
    \toprule
    Parameter & Model & \multicolumn{2}{c}{HR} & \multicolumn{2}{c}{NDCG} \\
    \midrule
    \multirow{4}{*}{Free embedding dimension} & & D = 16 & D=32 & D = 16 & D=32 \\
    & DiffNet & 0.3272 & 0.3337 & 0.1952 & 0.2075 \\
    & SVD++ & 0.2581 & 0.2727 & 0.1545 & 0.1632 \\
    & PinSage & 0.2952 & 0.2958 & 0.1758 & 0.1779 \\
    \midrule
    \multirow{4}{*}{Top-N ranking} & & N = 5  & N = 10 & N = 5  & N = 10 \\
    & DiffNet & 0.2276 & 0.3317 & 0.1779 & 0.2083 \\
    & SVD++& 0.1868 & 0.2631 & 0.1389 & 0.1511 \\
    & PinSage & 0.2099 & 0.3065 & 0.1536 & 0.1868 \\
    \bottomrule
    \end{tabular}
    \caption{Testing results of the DiffNet on the Yelp dataset after 500 epochs of training: we consider two types of hyperparameters (1) the number of dimension of the free embedding $\mathbf{P_a}$/$\mathbf{Q_i}$ while $N$ is fixed at 10 (2) the number of the top-N ranking for the predicted recommendation while $D$ is fixed at 16.}
    \label{tab:yelp_num_result}
\end{table*}

\section{Recommendation and Dissemination} \label{sec:dissemination}

The DiffNet enables a way to make recommendations by utilizing social information. The recommendation can be viewed as an establishment of new interaction between users and items. However, such a recommendation may worsen the information dissemination of the network, e.g. repetitively recommend contents from the same topic to a user or confine the recommendations for a user to a specific category. To avoid the recommendations with limited varieties, information dissemination capability of the graph formed from recommendation results can be considered to evaluate the recommendation performance. Thus, in this section, we first investigate the unsaturated level of information dissemination in real world graphs by comparing to other synthetic graphs with similar sizes in a connectivity optimization. Later, motivated by the unsaturated information dissemination capability, we propose a way to incorporate the dissemination capability of the network as an additional metric to recommendation tasks.

\subsection{Information Dissemination}
Information dissemination describes a characteristic of a graph in a stochastic process such as disease infection, news spreading, or data broadcasting. The prominent models to study these processes are the compartmental models which consider an ordinary differential equation (ODE) between states of nodes with interactions arising from the edges of the graph. As a simple variation, Susceptible-Infectious-Susceptible (SIS) models susceptible-to-infectious interaction via edges and probabilistic recovery to susceptible. Despite the complexity of the process, \citet{Wang03epidemicspreading} have discovered that information dissemination can be summarized by one parameter: the largest eigenvalue $\lambda_1$ of the adjacent matrix. Moreover, the largest eigenvalue also indicates whether a spreading would evolve into an epidemic or disappear over time \citet{Prakash2012Cascade}.

\begin{figure}
\centering
    \begin{subfigure}{\linewidth}
        \centering
        \includegraphics[width=0.7\linewidth,keepaspectratio]{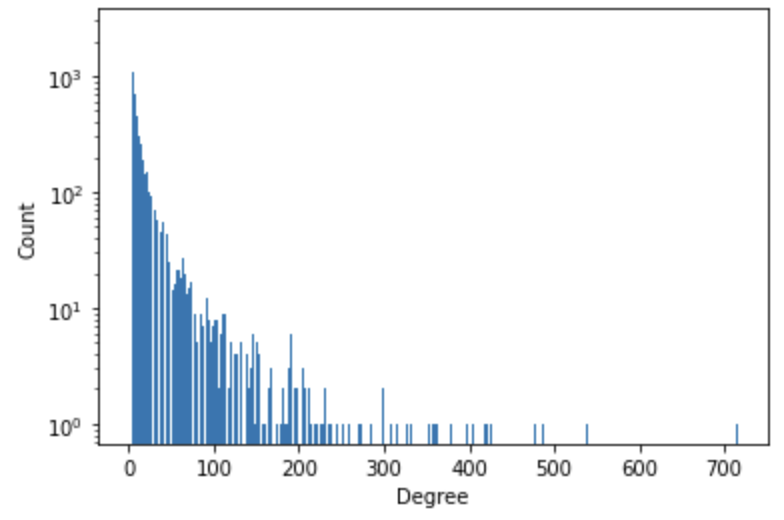}
        \caption{Node degree on Yelp}
    \end{subfigure} %
    \begin{subfigure}{\linewidth}
        \centering
        \includegraphics[width=0.7\linewidth,keepaspectratio]{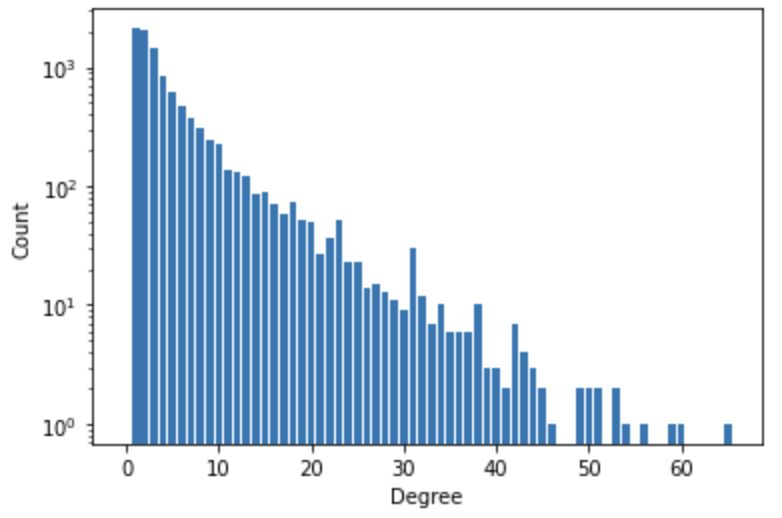}
        \caption{Node degree on CA-HEPTH}
     \end{subfigure} 
\caption{Histogram of node degree on Yelp (top) and CA-HEPTH (bottom) datasets. Note the y-axis is in logarithmic scale.}
\label{fig:dataset-degree}
\end{figure}

To this end, we study the selected datasets from the information dissemination perspective. In this report, we include two datasets: the Yelp social graph \citet{yelp_dataset} and the collaboration network of Arxiv High Energy Physics Theory (CA-HEPTH) \citet{Leskovec2007CollabNetworks}. To avoid an ill condition in the SIS model, we extract the largest connected subgraph from each dataset and remove all other connected components. The subgraphs contain 99.4\% and 87.5\% of all nodes. \autoref{tab:dataset-summary} shows the summary of dataset sizes and the information dissemination parameters $\lambda_1$. In addition, \autoref{fig:dataset-degree} sums up the degree distributions of the two datasets. Node degrees in both Yelp and CA-HEPTH generally follow the power-law distribution but Yelp dataset contains some nodes with extremely high degrees.

\begin{table}
    \centering
    \begin{tabular}{lrrr}
        \toprule
        Dataset & Nodes & Edges & $\lambda_1$ \\
        \midrule
        Yelp & 17138 & 129455 & 80.42 \\
        Stars Topology & 17138 & 129455 & 357.53 \\
        Next-K Neighborhood & 17138 & 129455 & 16.18 \\
        Erdős–Rényi Model & 17138 & 129455 & 16.16 \\
        Barabási–Albert Model & 17138 & 137040 & 43.54 \\
        \midrule
        CA-HEPTH & 8638 & 24827 & 31.03 \\
        Stars Topology & 8638 & 24827 & 156.40 \\
        Next-K Neighborhood & 8638 & 24827 & 6.00 \\
        Erdős–Rényi Model & 8638 & 24827 & 6.97 \\
        Barabási–Albert Model & 8638 & 25905 & 20.52 \\
        \bottomrule
    \end{tabular}
    \caption{Summary of datasets and their similar synthetics.}
    \label{tab:dataset-summary}
\end{table}

We compare these two real world graphs with 4 synthetic graphs with similar numbers of nodes and edges. Stars Topology refers to a graph where a few nodes connect to all other nodes. We connect the star's edges as many as we can to match the number of edges in the original graph. Given nodes labeled with consecutive numbers in a ring, Next-K Neighborhood links each node to its adjacent neighborhood such that the total number of edges match. Erdős–Rényi model uniformly samples all possible edges. Finally, Barabási–Albert model generates a graph with power-law degree distribution by sampling edges with preferential attachment. We generate these 4 synthetic graphs for each dataset size, totaling 8 generated graphs. Alongside the original graphs, we also list the summary of these synthetic graphs in \autoref{tab:dataset-summary}.

\begin{figure}
    \centering
    \includegraphics[width=0.7\linewidth,keepaspectratio]{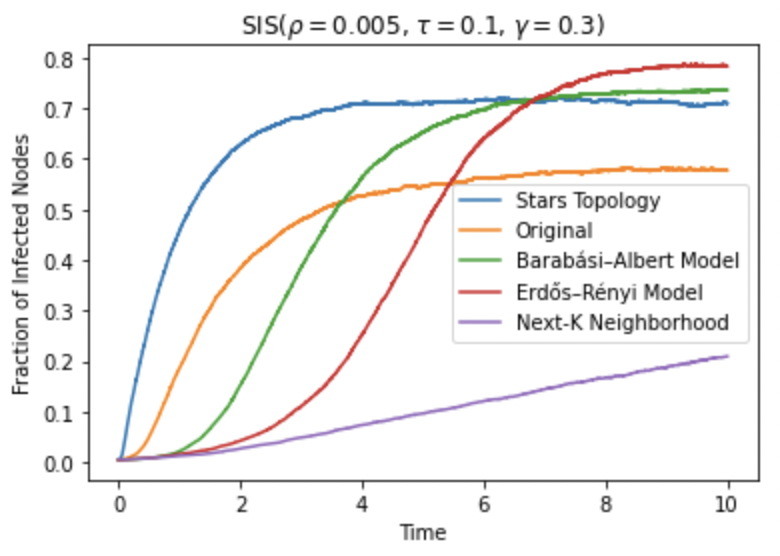}
    \includegraphics[width=0.7\linewidth,keepaspectratio]{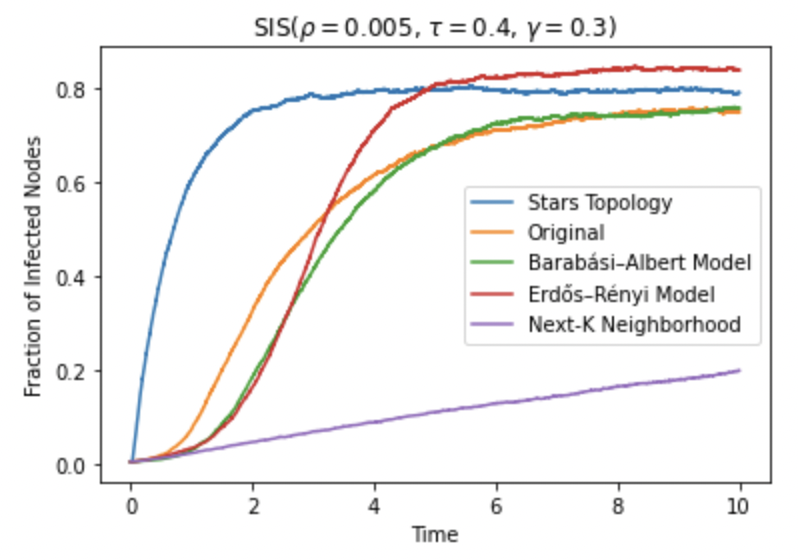}
    \caption{Fraction of infectious node over time in a SIS simulation of corresponding graph setting. Original graphs are Yelp (top) and CA-HEPTH (bottom). Lines of graphs in the legend are sorted by $\lambda_1$ without additional edge in descending order. The initial infectious fraction $\rho$, transmission rate $\tau$, and recovery rate $\gamma$ are denoted on the top of each plot.}
    \label{fig:infodis-sis}
\end{figure}

SIS simulations \citet{Miller2019Eon} in \autoref{fig:infodis-sis} empirically present the information dissemination process in all 10 graphs. Overall, the convergence rate of the infected fraction is consistent with the largest eigenvalues shown earlier: the higher $\lambda_1$ of the graph, the faster the convergence rate is initially. Stars Topology stands as the most information-disseminating graph. And this finding is consistent with the intuition: information is much easier to propagate in this setting since information from nodes with extremely high degrees can be quickly propagated to their numerous neighbors just one hop away. And the gap between the spreading in real world graphs and the artificial Stars Topology signifies some room for improvement for the real world network topology.

\begin{figure}
    \centering
    \includegraphics[width=0.7\linewidth,keepaspectratio]{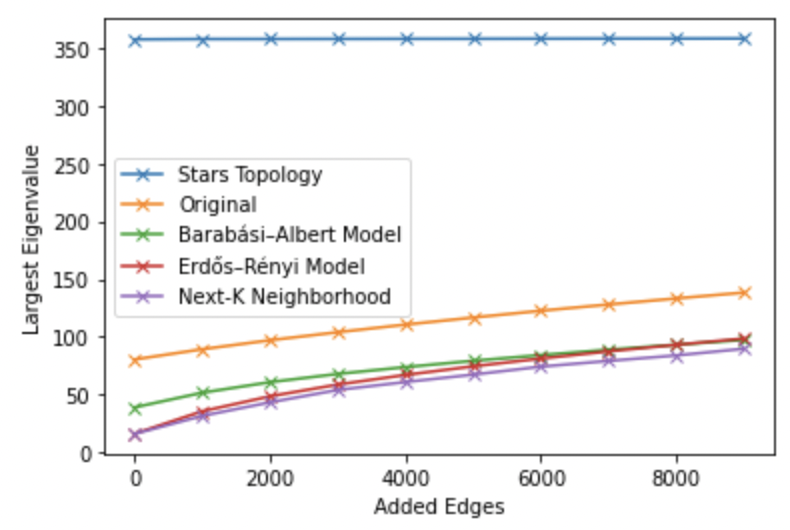}
    \includegraphics[width=0.7\linewidth,keepaspectratio]{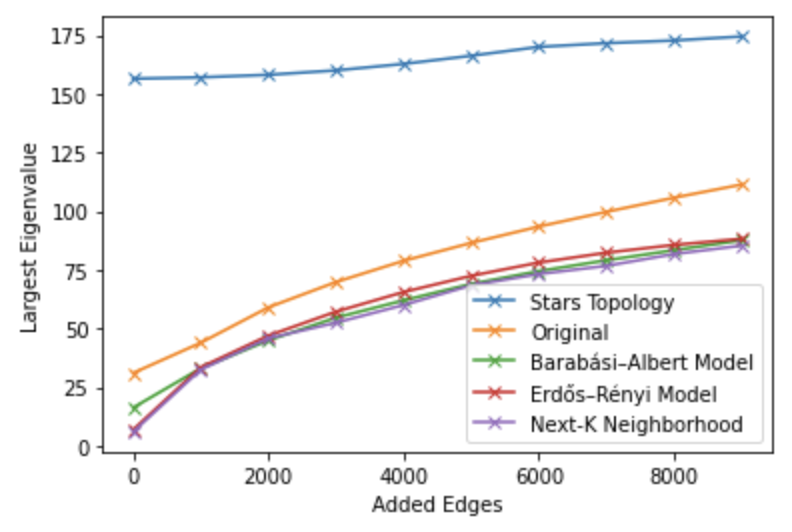}
    \caption{Largest eigenvalues $\lambda_1$ after adding numbers of edges suggested by gelling. Original graphs are Yelp (top) and CA-HEPTH (bottom). Lines of graphs in the legend are sorted by $\lambda_1$ without additional edge in descending order.}
    \label{fig:infodis-gelling}
\end{figure}

To further show that some of these graphs could have a wider spreading result, we replicate Gelling algorithm from \citet{Tong2012Gelling} to suggest edges that would improve information dissemination the most. We choose to implement the full $O(n^2)$ version of the proposed algorithm, given the dataset size is tractable. Figure \ref{fig:infodis-gelling} outlines increasing largest eigenvalues after adding some edges via the Gelling technique. When we add $10^4$ edges, the information dissemination capability of the graph from Yelp dataset doubles while that of CA-HEPTH dataset improves by approximately 4 times. Similar trends can be observed in other synthetic graphs as well, but only insignificant in Stars Topology.

\subsection{Dissemination-aware Training}
Now we see that there is a gap between ideally disseminating graph and existing ones. One can be tempted to substitute the recommender with dissemination optimizer; however, the most influential edge on dissemination is not guaranteed to be the best edge for recommendation task i.e. obtaining the highest HR or NDCG. Thus, our goal is to both accurately predict connections and improve dissemination in the graph to improve user's exploration on new information.

To recommend new user-item pairwise relation via a neural network model, the most common approach is using the cross entropy loss to quantify the difference between the predicted edge connecting probability and the true edge connection status. In this case, the edge of interest is the predicted user-item interaction that has not been seen yet. However, we later find that using mean squared error (MSE) loss can significantly improve the performance of the model in terms of HR/NDCG. And the objective of the dissemination-aware recommendation model is shown as following:
\begin{equation}\label{eqn:dis_aware_train_obj}
    \min_{\hat{\mathbf{y}}} \mathcal{L}(\hat{\mathbf{y}}; \mathbf{y}, \lambda) = \mathbb{E}_{a \sim \mathcal{A}, i \sim \mathcal{I}} \left[ \frac{1}{2} S_\alpha(a, i) (y_{a, i} - \hat{y}_{a, i})^2 \right],
\end{equation}

where $\lambda(a, i)$ is the dissemination score and $S_\alpha(u, i)$ is a dissemination factor between user $a$ in distribution $\mathcal{A}$ and item $i$ in distribution $\mathcal{I}$. In this study, the $S_\alpha(u, i)$ has the form:
\begin{equation}\label{eqn:dis_factor}
    S_\alpha(u, i) = \frac{1 + \lambda(a, i) e^{\alpha}}{1 + e^{\alpha}}.
\end{equation}

We assume a uniform distribution of user index. To mitigate the negative or unknown interactions dominating the training process, we construct an biased item distribution towards positive interactions similarly to other preceding evaluations \citet{wu2019neural, He2017}. Specifically, we use all the positive ratings from each user provided in the Yelp dataset as the true positive connections.But we only uniformly sample 8 out of the remaining possible user-item connections for training and assign them as the negative interactions (since those connections have not been observed in the real dataset). 
As for evaluating the recommendation performance, we select 1000 users randomly and prepare the positive/negative connections in the same manner to compute HR and NDCG.

Another important term introduced in the \autoref{eqn:dis_aware_train_obj} is the dissemination score $\lambda(a,i)$. To obtain the score for each possible user-item pairwise relation, before training, we first combine the user-user social network and the user-item interactions and derive an $M+N$ by $M+N$ adjacency matrix. Then the first eigenvector $\mathbf{z} \in \mathbb{R}^{M + N}$ of the aggregated adjacency matrix has been precomputed. Thus during the training process, we can use this eigenvector to compute the dissemination score $\lambda(a, i) = z[a] z[i]$.

Notice that the constant $\alpha \in \mathbb{R}$ in \autoref{eqn:dis_factor} is an additional hyperparameter. It can be interpreted as a interpolation weight between the original MSE loss without considering dissemination and its fully dissemination-aware version, when $\alpha \rightarrow -\infty$ and $\alpha \rightarrow \infty$ respectively. In our experiment, we have analyzed the effects of this hyperparameter and have varied $\alpha$ to generate the accuracy-dissemination Pareto front plot of the recommendation model.

\section{Experiment Results} \label{sec:experiments}
Besides the established metrics for evaluating the recommendation performance e.g. HR and NDCG, we also need to analyze the information dissemination capability of the graph formed after obtaining the recommendation results. And the dissemination performance is quantified by the largest eigenvalue $\lambda_1$ of the graph.

The original Yelp dataset has been split into train, validation, and test dataset in a way that each partition contains non-overlapping user-item interactions. And the test dataset has around 18000 records correspond to 10622 users and 11948 items, which is a sparse network that can be constructed quickly. The original $\lambda_1$ of the social network for the testing data is 70.47. To compare the information dissemination capability of the new networks, we have trained the dissemination-aware DiffNet for 500 epochs. Then we run the trained model on the testing dataset, obtain the recommendation results and expand the user-user social networks by adding user-item edges. For any user-item pair, if the predicted $\hat{r}_{ai}$ is higher than 0.5, a corresponding edge $e=(a,i)$ will be added to the original graph. 

\begin{table}
    \centering
    \begin{tabular}{ccccc}
        \toprule
        $\alpha$ & \# edges added & HR & NDCG & $\lambda_1$ \\
        \midrule
        3 & 18140 & 0.2077 & 0.1375 & 77.10\\
        2 & 17078 & 0.2113 & 0.1378 & 74.50\\
        1 & 19757 & 0.2376 & 0.1455 & 71.09\\
        0 & 16405 & 0.2653 & 0.1645 & 71.03\\
        -1 & 18955 & 0.2839 & 0.1733 & 70.42\\
        -2 & 16590 & 0.3098 & 0.1792 & 66.28\\
        -3 & 16792 & 0.3044 & 0.1811 & 63.55\\
        \midrule
        DiffNet & 17217 & 0.3317 & 0.2083 & 58.97\\
        SVD++ & 17685 & 0.2631 & 0.1531 & 58.63\\
        PinSage & 16338 & 0.3065 & 0.1868 & 60.83\\
        \bottomrule
    \end{tabular}
    \caption{Recommendation performance evaluated in HR and NDCG as well as the dissemination capability after introducing new connections based on the dissemination-aware DiffNet: for all tests, $D=16$ and $N=10$; last three rows correspond to results obtained from the original recommender systems without considering dissemination.}
    \label{tab:dis_aware_results}
\end{table}
\autoref{tab:dis_aware_results} summarizes the numerical results for the dissemination-aware DiffNet on the testing Yelp dataset with different weights $\alpha$. We have considered the HR/NDCG as well as the largest eigenvalue of the graph to evaluate the recommendation performance. And the performance of the original DiffNet without any dissemination term is denoted as $\alpha=-\infty$. It's not surprising to see that the recommendation accuracy has been deteriorated after introducing the dissemination term in the objective function while the largest eigenvalue increases significantly. For all levels of $\alpha$, the numbers of newly introduced edges i.e. predicted user-item connections are all around the same level. However, the recommendation accuracy and the dissemination capability achieved from different settings are quite different. This indicates that the final topologies of the user-item graph are different. And the final dissemination scores in some cases are worse than the original social networks without any recommendation e.g. when $\alpha \leq -1$. This is due to the fact that most of the predicted recommendations connect new item to existing users as an end node, which has a low nodal degree in the resulting graph.

\begin{figure}
\centering
    \begin{subfigure}{\linewidth}
        \centering
        \includegraphics[width=0.7\linewidth,keepaspectratio]{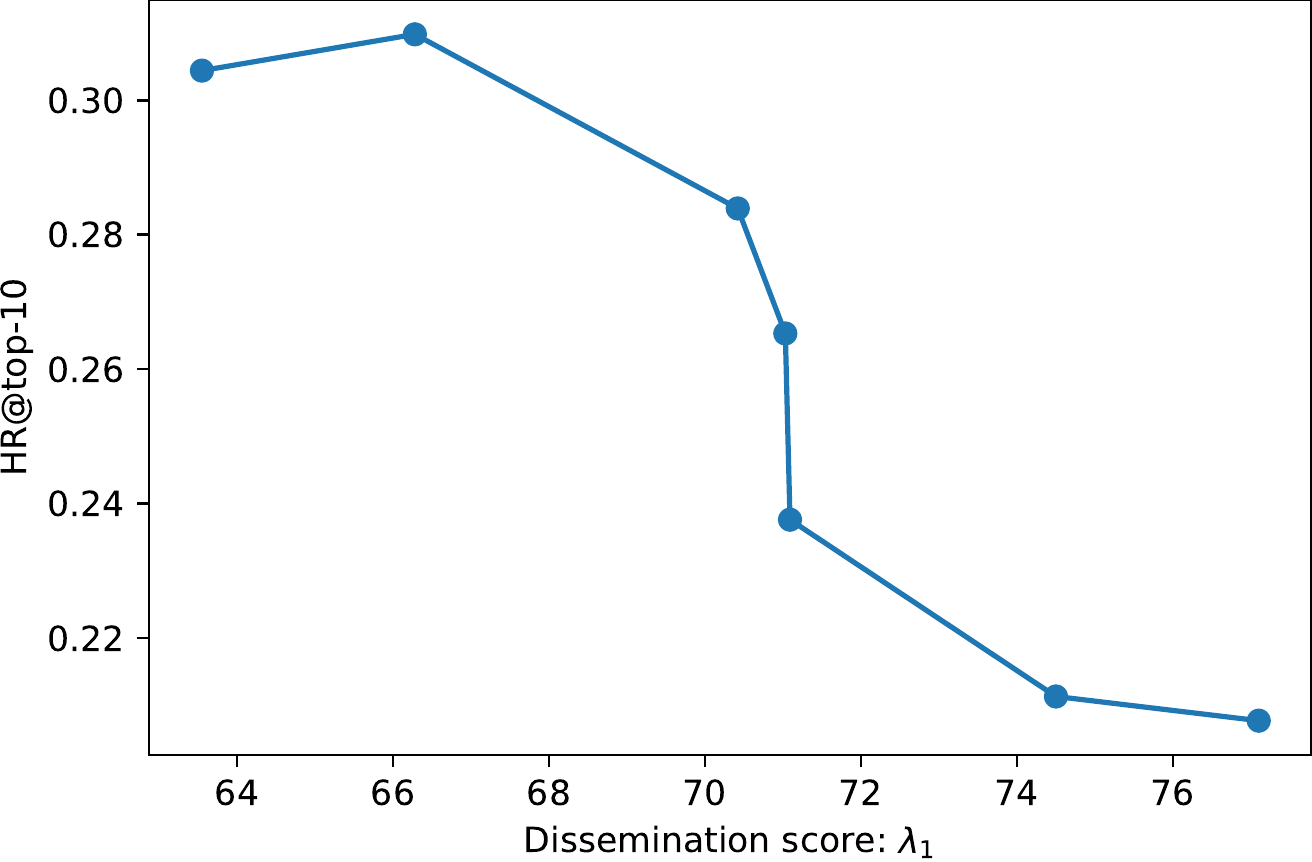}
        \caption{HR versus $\lambda_1$}
    \end{subfigure} %
    \begin{subfigure}{\linewidth}
        \centering
        \includegraphics[width=0.7\linewidth,keepaspectratio]{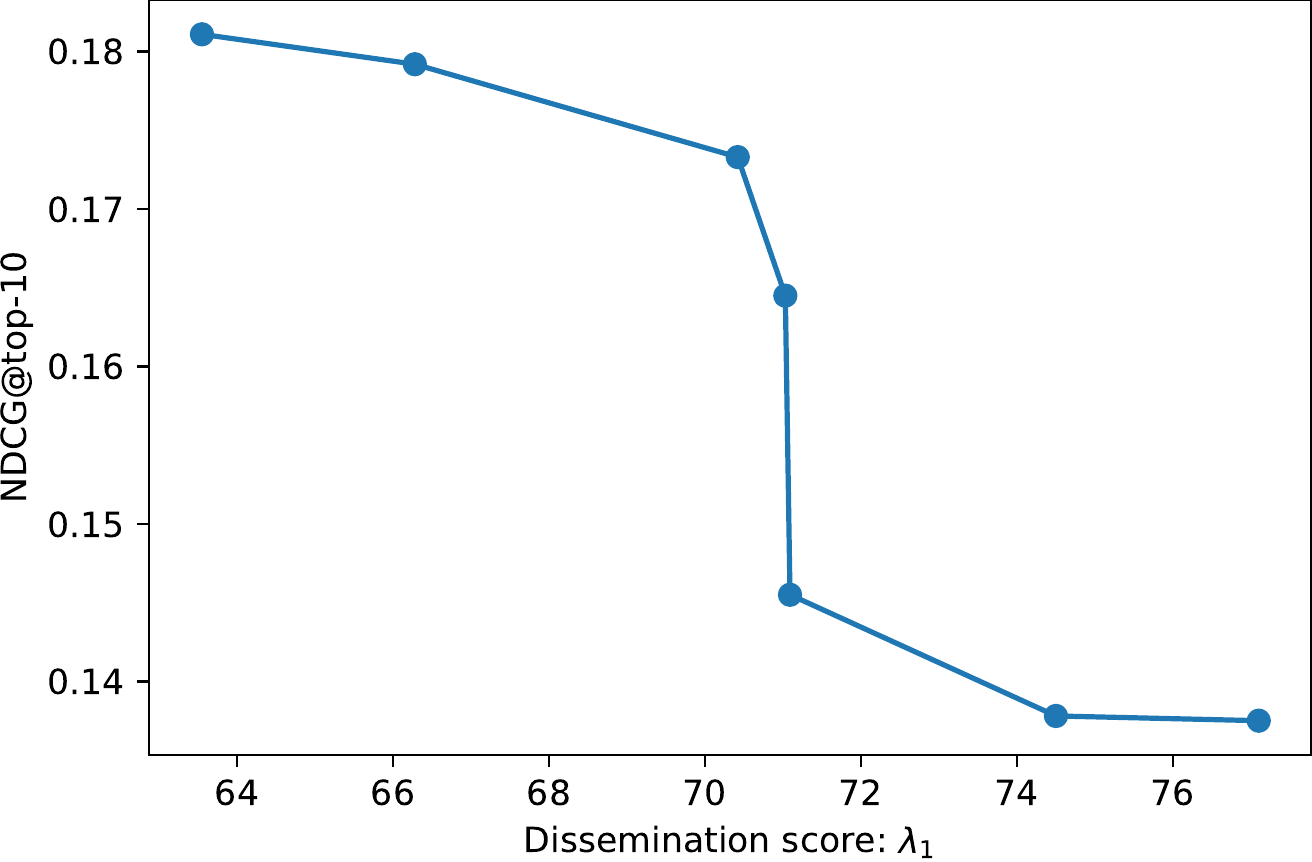}
        \caption{NDCG versus $\lambda_1$}
     \end{subfigure} 
\caption{Visualization of trade-off between the recommendation accuracy and the information dissemination capability of the network after realizing recommendation results.}
\label{fig:pareto_results}
\end{figure}

Based on the numerical results in \autoref{tab:dis_aware_results}, we can see a clear trade-off between the recommendation accuracy and the information dissemination capability induced by the recommendation algorithm. To better comprehend the trad-off, \autoref{fig:pareto_results} illustrates the changes in NR/NDCG and the $\lambda_1$ obtained from the new user-item graph.
For both the NR and NDCG, the numerical results decrease with the increasing dissemination score $\lambda_1$. And this observation is consistent with the objective function since larger $\alpha$ makes the recommendation model focus more on establishing edges to maximize the information dissemination, while smaller $\alpha$ leads to a model similar to the vanilla DiffNet. From \autoref{fig:pareto_results}, we can see that the decreasing trend of the recommendation accuracy is not linear and there is a sharp change when the $\alpha$ is set to be approximately 0. This could be the result of expressing the dissemination factor as an exponential term. And the plot suggests that an $\alpha$ between 0 and -1 is a good choice for such a recommendation task on the Yelp dataset. Since the recommendation accuracy is acceptable (even higher than the SVD++ results) while the information dissemination capability has been maintained at a high level.

\section{Related Works} \label{sec:related_works}

Graph neural networks (GNNs) recently find their applications in recommender systems. \citet{Wu2020GraphNN} compiles an extensive survey of these approaches based on different GNN structure (GCN \citet{FU202192}, GraphSage \citet{Hamilton2017GraphSage}, GAT \citet{velickovic2018graph}, and GGGN \citet{li2016gated}), temporal dependent (general recommendation, sequential recommendation), and considered information (user-item interactions, social network, knowledge graph). In terms of this taxonomy, DiffNet's concept of influence diffusion is closely related to GCN approach. DiffNet performs a general recommendation and so assumes that recommendation is invariant to time. It also relies on social network to help learning user-item interactions.

Apart from dissemination, many other works enhance recommendation task with auxiliary properties to solve specific problems \citet{Ricci2010Recommender}. Many objectives align well with our interest in explorative recommendations. For example, \textit{serendipity} measures the surprising degree of successful recommendations which encourage recommendations out of the typical interactions. \textit{Novelty} and \textit{diversity} are also well-known desirable properties in a recommender systems which determine the distinction among the recommendations. \citet{Hurley2011RecDiversity} explains various approaches to enhance existing frameworks towards these metrics. Instead of directly rewarding a diverse recommendation list, our algorithm globally focuses on the effect of successful recommendation on the user's future exploration, reflected in the dissemination term. 


\section{Conclusion}
In this project, we have studied existing recommender systems utilizing graph information such as user-user and user-item interactions. To improve the effectiveness of the recommendations in terms of both the accuracy and user's exploration, we have considered the information dissemination of the graph as an aid in the recommendation algorithm. Case studies based on the Yelp dataset have successfully demonstrated the advantage of the proposed framework: comparing to baseline models such as SVD++ and PinSage, the recommendation accuracy has been maintained at desirable levels while the information dissemination capability of the formed user-item graph has been optimized. As for future directions, more comprehensive diffusion process for the network can be utilized: not only the user embedding can be evolved through the information propagation process but also the item embedding can be iteratively refined.








\bibliographystyle{ACM-Reference-Format}
\bibliography{Project_midterm_report}


\begin{thebibliography}{25}


\ifx \showCODEN    \undefined \def \showCODEN     #1{\unskip}     \fi
\ifx \showDOI      \undefined \def \showDOI       #1{#1}\fi
\ifx \showISBNx    \undefined \def \showISBNx     #1{\unskip}     \fi
\ifx \showISBNxiii \undefined \def \showISBNxiii  #1{\unskip}     \fi
\ifx \showISSN     \undefined \def \showISSN      #1{\unskip}     \fi
\ifx \showLCCN     \undefined \def \showLCCN      #1{\unskip}     \fi
\ifx \shownote     \undefined \def \shownote      #1{#1}          \fi
\ifx \showarticletitle \undefined \def \showarticletitle #1{#1}   \fi
\ifx \showURL      \undefined \def \showURL       {\relax}        \fi
\providecommand\bibfield[2]{#2}
\providecommand\bibinfo[2]{#2}
\providecommand\natexlab[1]{#1}
\providecommand\showeprint[2][]{arXiv:#2}

\bibitem[\protect\citeauthoryear{??}{yel}{[n.d.]}]%
        {yelp_dataset}
 \bibinfo{year}{[n.d.]}\natexlab{}.
\newblock \bibinfo{title}{Yelp Open Dataset}.
\newblock
\newblock
\urldef\tempurl%
\url{https://www.yelp.com/dataset}
\showURL{%
\tempurl}


\bibitem[\protect\citeauthoryear{Fan, Ma, Li, He, Zhao, Tang, and Yin}{Fan
  et~al\mbox{.}}{2019}]%
        {Fan2019}
\bibfield{author}{\bibinfo{person}{Wenqi Fan}, \bibinfo{person}{Yao Ma},
  \bibinfo{person}{Qing Li}, \bibinfo{person}{Yuan He}, \bibinfo{person}{Eric
  Zhao}, \bibinfo{person}{Jiliang Tang}, {and} \bibinfo{person}{Dawei Yin}.}
  \bibinfo{year}{2019}\natexlab{}.
\newblock \showarticletitle{Graph Neural Networks for Social Recommendation}.
  In \bibinfo{booktitle}{\emph{The World Wide Web Conference}} (San Francisco,
  CA, USA) \emph{(\bibinfo{series}{WWW '19})}. \bibinfo{publisher}{Association
  for Computing Machinery}, \bibinfo{address}{New York, NY, USA},
  \bibinfo{pages}{417–426}.
\newblock
\showISBNx{9781450366748}
\urldef\tempurl%
\url{https://doi.org/10.1145/3308558.3313488}
\showDOI{\tempurl}


\bibitem[\protect\citeauthoryear{Fu, Liu, Zhang, Zhou, and Tao}{Fu
  et~al\mbox{.}}{2021}]%
        {FU202192}
\bibfield{author}{\bibinfo{person}{Sichao Fu}, \bibinfo{person}{Weifeng Liu},
  \bibinfo{person}{Kai Zhang}, \bibinfo{person}{Yicong Zhou}, {and}
  \bibinfo{person}{Dapeng Tao}.} \bibinfo{year}{2021}\natexlab{}.
\newblock \showarticletitle{Semi-supervised classification by graph p-Laplacian
  convolutional networks}.
\newblock \bibinfo{journal}{\emph{Information Sciences}}  \bibinfo{volume}{560}
  (\bibinfo{year}{2021}), \bibinfo{pages}{92--106}.
\newblock
\showISSN{0020-0255}
\urldef\tempurl%
\url{https://doi.org/10.1016/j.ins.2021.01.075}
\showDOI{\tempurl}


\bibitem[\protect\citeauthoryear{Guo, Zhang, and Yorke-Smith}{Guo
  et~al\mbox{.}}{2015a}]%
        {Guo2015}
\bibfield{author}{\bibinfo{person}{Guibing Guo}, \bibinfo{person}{Jie Zhang},
  {and} \bibinfo{person}{Neil Yorke-Smith}.} \bibinfo{year}{2015}\natexlab{a}.
\newblock \showarticletitle{TrustSVD: Collaborative Filtering with Both the
  Explicit and Implicit Influence of User Trust and of Item Ratings}. In
  \bibinfo{booktitle}{\emph{Proceedings of the Twenty-Ninth AAAI Conference on
  Artificial Intelligence}} (Austin, Texas) \emph{(\bibinfo{series}{AAAI'15})}.
  \bibinfo{publisher}{AAAI Press}, \bibinfo{pages}{123–129}.
\newblock
\showISBNx{0262511290}


\bibitem[\protect\citeauthoryear{Guo, Zhang, and Yorke-Smith}{Guo
  et~al\mbox{.}}{2015b}]%
        {guo2015trustsvd}
\bibfield{author}{\bibinfo{person}{Guibing Guo}, \bibinfo{person}{Jie Zhang},
  {and} \bibinfo{person}{Neil Yorke-Smith}.} \bibinfo{year}{2015}\natexlab{b}.
\newblock \showarticletitle{Trustsvd: Collaborative filtering with both the
  explicit and implicit influence of user trust and of item ratings}. In
  \bibinfo{booktitle}{\emph{Proceedings of the AAAI Conference on Artificial
  Intelligence}}, Vol.~\bibinfo{volume}{29}.
\newblock


\bibitem[\protect\citeauthoryear{Guo, Tang, Ye, Li, and He}{Guo
  et~al\mbox{.}}{2017}]%
        {Guo2017}
\bibfield{author}{\bibinfo{person}{Huifeng Guo}, \bibinfo{person}{Ruiming
  Tang}, \bibinfo{person}{Yunming Ye}, \bibinfo{person}{Zhenguo Li}, {and}
  \bibinfo{person}{Xiuqiang He}.} \bibinfo{year}{2017}\natexlab{}.
\newblock \showarticletitle{DeepFM: A Factorization-Machine Based Neural
  Network for CTR Prediction}. In \bibinfo{booktitle}{\emph{Proceedings of the
  26th International Joint Conference on Artificial Intelligence}} (Melbourne,
  Australia) \emph{(\bibinfo{series}{IJCAI'17})}. \bibinfo{publisher}{AAAI
  Press}, \bibinfo{pages}{1725–1731}.
\newblock
\showISBNx{9780999241103}


\bibitem[\protect\citeauthoryear{Hamilton, Ying, and Leskovec}{Hamilton
  et~al\mbox{.}}{2017}]%
        {Hamilton2017GraphSage}
\bibfield{author}{\bibinfo{person}{William~L. Hamilton}, \bibinfo{person}{Rex
  Ying}, {and} \bibinfo{person}{Jure Leskovec}.}
  \bibinfo{year}{2017}\natexlab{}.
\newblock \showarticletitle{Inductive Representation Learning on Large Graphs}.
  In \bibinfo{booktitle}{\emph{Proceedings of the 31st International Conference
  on Neural Information Processing Systems}} (Long Beach, California, USA)
  \emph{(\bibinfo{series}{NIPS'17})}. \bibinfo{publisher}{Curran Associates
  Inc.}, \bibinfo{address}{Red Hook, NY, USA}, \bibinfo{pages}{1025–1035}.
\newblock
\showISBNx{9781510860964}


\bibitem[\protect\citeauthoryear{He, Liao, Zhang, Nie, Hu, and Chua}{He
  et~al\mbox{.}}{2017}]%
        {He2017}
\bibfield{author}{\bibinfo{person}{Xiangnan He}, \bibinfo{person}{Lizi Liao},
  \bibinfo{person}{Hanwang Zhang}, \bibinfo{person}{Liqiang Nie},
  \bibinfo{person}{Xia Hu}, {and} \bibinfo{person}{Tat-Seng Chua}.}
  \bibinfo{year}{2017}\natexlab{}.
\newblock \showarticletitle{Neural Collaborative Filtering}. In
  \bibinfo{booktitle}{\emph{Proceedings of the 26th International Conference on
  World Wide Web}} (Perth, Australia) \emph{(\bibinfo{series}{WWW '17})}.
  \bibinfo{publisher}{International World Wide Web Conferences Steering
  Committee}, \bibinfo{address}{Republic and Canton of Geneva, CHE},
  \bibinfo{pages}{173–182}.
\newblock
\showISBNx{9781450349130}
\urldef\tempurl%
\url{https://doi.org/10.1145/3038912.3052569}
\showDOI{\tempurl}


\bibitem[\protect\citeauthoryear{Hurley and Zhang}{Hurley and Zhang}{2011}]%
        {Hurley2011RecDiversity}
\bibfield{author}{\bibinfo{person}{Neil Hurley} {and} \bibinfo{person}{Mi
  Zhang}.} \bibinfo{year}{2011}\natexlab{}.
\newblock \showarticletitle{Novelty and Diversity in Top-N Recommendation --
  Analysis and Evaluation}.
\newblock \bibinfo{journal}{\emph{ACM Trans. Internet Technol.}}
  \bibinfo{volume}{10}, \bibinfo{number}{4}, Article \bibinfo{articleno}{14}
  (\bibinfo{date}{March} \bibinfo{year}{2011}), \bibinfo{numpages}{30}~pages.
\newblock
\showISSN{1533-5399}
\urldef\tempurl%
\url{https://doi.org/10.1145/1944339.1944341}
\showDOI{\tempurl}


\bibitem[\protect\citeauthoryear{{Jiang}, {Cui}, {Wang}, {Zhu}, and
  {Yang}}{{Jiang} et~al\mbox{.}}{2014}]%
        {Jiang2014}
\bibfield{author}{\bibinfo{person}{M. {Jiang}}, \bibinfo{person}{P. {Cui}},
  \bibinfo{person}{F. {Wang}}, \bibinfo{person}{W. {Zhu}}, {and}
  \bibinfo{person}{S. {Yang}}.} \bibinfo{year}{2014}\natexlab{}.
\newblock \showarticletitle{Scalable Recommendation with Social Contextual
  Information}.
\newblock \bibinfo{journal}{\emph{IEEE Transactions on Knowledge and Data
  Engineering}} \bibinfo{volume}{26}, \bibinfo{number}{11}
  (\bibinfo{year}{2014}), \bibinfo{pages}{2789--2802}.
\newblock
\urldef\tempurl%
\url{https://doi.org/10.1109/TKDE.2014.2300487}
\showDOI{\tempurl}


\bibitem[\protect\citeauthoryear{Konstas, Stathopoulos, and Jose}{Konstas
  et~al\mbox{.}}{2009}]%
        {Konstas2009}
\bibfield{author}{\bibinfo{person}{Ioannis Konstas}, \bibinfo{person}{Vassilios
  Stathopoulos}, {and} \bibinfo{person}{Joemon~M. Jose}.}
  \bibinfo{year}{2009}\natexlab{}.
\newblock \showarticletitle{On Social Networks and Collaborative
  Recommendation}. In \bibinfo{booktitle}{\emph{Proceedings of the 32nd
  International ACM SIGIR Conference on Research and Development in Information
  Retrieval}} (Boston, MA, USA) \emph{(\bibinfo{series}{SIGIR '09})}.
  \bibinfo{publisher}{Association for Computing Machinery},
  \bibinfo{address}{New York, NY, USA}, \bibinfo{pages}{195–202}.
\newblock
\showISBNx{9781605584836}
\urldef\tempurl%
\url{https://doi.org/10.1145/1571941.1571977}
\showDOI{\tempurl}


\bibitem[\protect\citeauthoryear{Leskovec, Kleinberg, and Faloutsos}{Leskovec
  et~al\mbox{.}}{2007}]%
        {Leskovec2007CollabNetworks}
\bibfield{author}{\bibinfo{person}{Jure Leskovec}, \bibinfo{person}{Jon
  Kleinberg}, {and} \bibinfo{person}{Christos Faloutsos}.}
  \bibinfo{year}{2007}\natexlab{}.
\newblock \showarticletitle{Graph Evolution: Densification and Shrinking
  Diameters}.
\newblock \bibinfo{journal}{\emph{ACM Trans. Knowl. Discov. Data}}
  \bibinfo{volume}{1}, \bibinfo{number}{1} (\bibinfo{date}{March}
  \bibinfo{year}{2007}), \bibinfo{pages}{2–es}.
\newblock
\showISSN{1556-4681}
\urldef\tempurl%
\url{https://doi.org/10.1145/1217299.1217301}
\showDOI{\tempurl}


\bibitem[\protect\citeauthoryear{Li, Zemel, Brockschmidt, and Tarlow}{Li
  et~al\mbox{.}}{2016}]%
        {li2016gated}
\bibfield{author}{\bibinfo{person}{Yujia Li}, \bibinfo{person}{Richard Zemel},
  \bibinfo{person}{Marc Brockschmidt}, {and} \bibinfo{person}{Daniel Tarlow}.}
  \bibinfo{year}{2016}\natexlab{}.
\newblock \showarticletitle{Gated Graph Sequence Neural Networks}. In
  \bibinfo{booktitle}{\emph{Proceedings of ICLR'16}
  (\bibinfo{edition}{proceedings of iclr'16} ed.)}.
\newblock
\urldef\tempurl%
\url{https://www.microsoft.com/en-us/research/publication/gated-graph-sequence-neural-networks/}
\showURL{%
\tempurl}


\bibitem[\protect\citeauthoryear{Miller and Ting}{Miller and Ting}{2019}]%
        {Miller2019Eon}
\bibfield{author}{\bibinfo{person}{Joel~C. Miller} {and} \bibinfo{person}{Tony
  Ting}.} \bibinfo{year}{2019}\natexlab{}.
\newblock \showarticletitle{EoN (Epidemics on Networks): a fast, flexible
  Python package for simulation, analytic approximation, and analysis of
  epidemics on networks}.
\newblock \bibinfo{journal}{\emph{Journal of Open Source Software}}
  \bibinfo{volume}{4}, \bibinfo{number}{44} (\bibinfo{year}{2019}),
  \bibinfo{pages}{1731}.
\newblock
\urldef\tempurl%
\url{https://doi.org/10.21105/joss.01731}
\showDOI{\tempurl}


\bibitem[\protect\citeauthoryear{Prakash, Chakrabarti, Valler, Faloutsos, and
  Faloutsos}{Prakash et~al\mbox{.}}{2012}]%
        {Prakash2012Cascade}
\bibfield{author}{\bibinfo{person}{B.~Aditya Prakash},
  \bibinfo{person}{Deepayan Chakrabarti}, \bibinfo{person}{Nicholas~C. Valler},
  \bibinfo{person}{Michalis Faloutsos}, {and} \bibinfo{person}{Christos
  Faloutsos}.} \bibinfo{year}{2012}\natexlab{}.
\newblock \showarticletitle{Threshold Conditions for Arbitrary Cascade Models
  on Arbitrary Networks}.
\newblock \bibinfo{journal}{\emph{Knowl. Inf. Syst.}} \bibinfo{volume}{33},
  \bibinfo{number}{3} (\bibinfo{date}{Dec.} \bibinfo{year}{2012}),
  \bibinfo{pages}{549–575}.
\newblock
\showISSN{0219-1377}
\urldef\tempurl%
\url{https://doi.org/10.1007/s10115-012-0520-y}
\showDOI{\tempurl}


\bibitem[\protect\citeauthoryear{Rendle}{Rendle}{2012}]%
        {Rendle2012}
\bibfield{author}{\bibinfo{person}{Steffen Rendle}.}
  \bibinfo{year}{2012}\natexlab{}.
\newblock \showarticletitle{Factorization Machines with LibFM}.
\newblock \bibinfo{journal}{\emph{ACM Trans. Intell. Syst. Technol.}}
  \bibinfo{volume}{3}, \bibinfo{number}{3}, Article \bibinfo{articleno}{57}
  (\bibinfo{date}{May} \bibinfo{year}{2012}), \bibinfo{numpages}{22}~pages.
\newblock
\showISSN{2157-6904}
\urldef\tempurl%
\url{https://doi.org/10.1145/2168752.2168771}
\showDOI{\tempurl}


\bibitem[\protect\citeauthoryear{Ricci, Rokach, Shapira, and Kantor}{Ricci
  et~al\mbox{.}}{2010}]%
        {Ricci2010Recommender}
\bibfield{author}{\bibinfo{person}{Francesco Ricci}, \bibinfo{person}{Lior
  Rokach}, \bibinfo{person}{Bracha Shapira}, {and} \bibinfo{person}{Paul~B.
  Kantor}.} \bibinfo{year}{2010}\natexlab{}.
\newblock \bibinfo{booktitle}{\emph{Recommender Systems Handbook}
  (\bibinfo{edition}{1st} ed.)}.
\newblock \bibinfo{publisher}{Springer-Verlag}, \bibinfo{address}{Berlin,
  Heidelberg}.
\newblock
\showISBNx{0387858199}


\bibitem[\protect\citeauthoryear{Tang, Hu, and Liu}{Tang et~al\mbox{.}}{2013}]%
        {tang2013social}
\bibfield{author}{\bibinfo{person}{Jiliang Tang}, \bibinfo{person}{Xia Hu},
  {and} \bibinfo{person}{Huan Liu}.} \bibinfo{year}{2013}\natexlab{}.
\newblock \showarticletitle{Social recommendation: a review}.
\newblock \bibinfo{journal}{\emph{Social Network Analysis and Mining}}
  \bibinfo{volume}{3}, \bibinfo{number}{4} (\bibinfo{year}{2013}),
  \bibinfo{pages}{1113--1133}.
\newblock


\bibitem[\protect\citeauthoryear{Tong, Prakash, Eliassi-Rad, Faloutsos, and
  Faloutsos}{Tong et~al\mbox{.}}{2012}]%
        {Tong2012Gelling}
\bibfield{author}{\bibinfo{person}{Hanghang Tong}, \bibinfo{person}{B.~Aditya
  Prakash}, \bibinfo{person}{Tina Eliassi-Rad}, \bibinfo{person}{Michalis
  Faloutsos}, {and} \bibinfo{person}{Christos Faloutsos}.}
  \bibinfo{year}{2012}\natexlab{}.
\newblock \showarticletitle{Gelling, and Melting, Large Graphs by Edge
  Manipulation}. In \bibinfo{booktitle}{\emph{Proceedings of the 21st ACM
  International Conference on Information and Knowledge Management}} (Maui,
  Hawaii, USA) \emph{(\bibinfo{series}{CIKM '12})}.
  \bibinfo{publisher}{Association for Computing Machinery},
  \bibinfo{address}{New York, NY, USA}, \bibinfo{pages}{245–254}.
\newblock
\showISBNx{9781450311564}
\urldef\tempurl%
\url{https://doi.org/10.1145/2396761.2396795}
\showDOI{\tempurl}


\bibitem[\protect\citeauthoryear{Veličković, Cucurull, Casanova, Romero,
  Liò, and Bengio}{Veličković et~al\mbox{.}}{2018}]%
        {velickovic2018graph}
\bibfield{author}{\bibinfo{person}{Petar Veličković},
  \bibinfo{person}{Guillem Cucurull}, \bibinfo{person}{Arantxa Casanova},
  \bibinfo{person}{Adriana Romero}, \bibinfo{person}{Pietro Liò}, {and}
  \bibinfo{person}{Yoshua Bengio}.} \bibinfo{year}{2018}\natexlab{}.
\newblock \showarticletitle{Graph Attention Networks}. In
  \bibinfo{booktitle}{\emph{International Conference on Learning
  Representations}}.
\newblock
\urldef\tempurl%
\url{https://openreview.net/forum?id=rJXMpikCZ}
\showURL{%
\tempurl}


\bibitem[\protect\citeauthoryear{Wang, Chakrabarti, Wang, and Faloutsos}{Wang
  et~al\mbox{.}}{2003}]%
        {Wang03epidemicspreading}
\bibfield{author}{\bibinfo{person}{Yang Wang}, \bibinfo{person}{Deepayan
  Chakrabarti}, \bibinfo{person}{Chenxi Wang}, {and} \bibinfo{person}{Christos
  Faloutsos}.} \bibinfo{year}{2003}\natexlab{}.
\newblock \showarticletitle{Epidemic Spreading in Real Networks: An Eigenvalue
  Viewpoint}. In \bibinfo{booktitle}{\emph{In SRDS}}. \bibinfo{pages}{25--34}.
\newblock


\bibitem[\protect\citeauthoryear{Wu, Sun, Fu, Hong, Wang, and Wang}{Wu
  et~al\mbox{.}}{2019}]%
        {wu2019neural}
\bibfield{author}{\bibinfo{person}{Le Wu}, \bibinfo{person}{Peijie Sun},
  \bibinfo{person}{Yanjie Fu}, \bibinfo{person}{Richang Hong},
  \bibinfo{person}{Xiting Wang}, {and} \bibinfo{person}{Meng Wang}.}
  \bibinfo{year}{2019}\natexlab{}.
\newblock \showarticletitle{A neural influence diffusion model for social
  recommendation}. In \bibinfo{booktitle}{\emph{Proceedings of the 42nd
  international ACM SIGIR conference on research and development in information
  retrieval}}. \bibinfo{pages}{235--244}.
\newblock


\bibitem[\protect\citeauthoryear{Wu, Zhang, Sun, and Cui}{Wu
  et~al\mbox{.}}{2020}]%
        {Wu2020GraphNN}
\bibfield{author}{\bibinfo{person}{S. Wu}, \bibinfo{person}{Wentao Zhang},
  \bibinfo{person}{Fei Sun}, {and} \bibinfo{person}{Bin Cui}.}
  \bibinfo{year}{2020}\natexlab{}.
\newblock \showarticletitle{Graph Neural Networks in Recommender Systems: A
  Survey}.
\newblock \bibinfo{journal}{\emph{ArXiv}}  \bibinfo{volume}{abs/2011.02260}
  (\bibinfo{year}{2020}).
\newblock


\bibitem[\protect\citeauthoryear{Yang, Steck, and Liu}{Yang
  et~al\mbox{.}}{2012}]%
        {Yang2012}
\bibfield{author}{\bibinfo{person}{Xiwang Yang}, \bibinfo{person}{Harald
  Steck}, {and} \bibinfo{person}{Yong Liu}.} \bibinfo{year}{2012}\natexlab{}.
\newblock \showarticletitle{Circle-Based Recommendation in Online Social
  Networks}. In \bibinfo{booktitle}{\emph{Proceedings of the 18th ACM SIGKDD
  International Conference on Knowledge Discovery and Data Mining}} (Beijing,
  China) \emph{(\bibinfo{series}{KDD '12})}. \bibinfo{publisher}{Association
  for Computing Machinery}, \bibinfo{address}{New York, NY, USA},
  \bibinfo{pages}{1267–1275}.
\newblock
\showISBNx{9781450314626}
\urldef\tempurl%
\url{https://doi.org/10.1145/2339530.2339728}
\showDOI{\tempurl}


\bibitem[\protect\citeauthoryear{Ying, He, Chen, Eksombatchai, Hamilton, and
  Leskovec}{Ying et~al\mbox{.}}{2018}]%
        {ying2018graph}
\bibfield{author}{\bibinfo{person}{Rex Ying}, \bibinfo{person}{Ruining He},
  \bibinfo{person}{Kaifeng Chen}, \bibinfo{person}{Pong Eksombatchai},
  \bibinfo{person}{William~L Hamilton}, {and} \bibinfo{person}{Jure Leskovec}.}
  \bibinfo{year}{2018}\natexlab{}.
\newblock \showarticletitle{Graph convolutional neural networks for web-scale
  recommender systems}. In \bibinfo{booktitle}{\emph{Proceedings of the 24th
  ACM SIGKDD International Conference on Knowledge Discovery \& Data Mining}}.
  \bibinfo{pages}{974--983}.
\newblock


\end{thebibliography}

\appendix

\end{document}